\documentclass{appolb}
\usepackage{graphicx}
\usepackage{amsmath}
\usepackage[labelsep=period]{caption}
\renewcommand{\thetable}{\Roman{table}}
\providecommand{\keywords}[1]
{
  \small	
  \textbf{\textit{Keywords---}} #1
}
\begin{document}
\title{Two-photon decay of para-positronium\\ within a composite approach%
\thanks{Based on the presentation of M.P. at the $5^{\text{th}}$ Jagiellonian Symposium on `Advances in Particle Physics and Medicine', June 29 - July 7 2024, Cracow, Poland.\\
$^a$e-mail: milena.piotrowska@ujk.edu.pl}%
}
\author{M. Piotrowska$^{1,a}$, F. Giacosa$^{1,2}$
\address{$^1$ Institute of Physics, Jan Kochanowski University,\\ \textit{ul. Uniwersytecka 7, 25-406 Kielce, Poland.}\\
$^{2}$ Institute for Theoretical Physics, J. W. Goethe University, \\ \textit{Max-von-Laue-Str. 1, 60438 Frankfurt, Germany.}\\}
%\\[3mm]
%{Third Author % of different affiliation
%\address{affiliation}
%}
%\\[3mm]
%the Name(s) of other Author(s)
%\address{affiliation}
}
\maketitle
\begin{abstract}
The decay of the para-positronium into two photons is studied in the framework of a composite Quantum Field Theoretical approach. This amounts to the evaluation of the electron-positron dressing, the Weinberg compositeness condition for the positronium, and the triangle-shaped diagram with virtual electrons circulating in it, leading to the final two-photon state. 
An important role is played by the positronium-electron-positron vertex, which is linked to the wave function of the para-positronium. We show how possible choices for the vertex function affect the $\gamma \gamma$ decay rate. Outlooks to other decay channels and to other positronia are presented.

\end{abstract}
  
\section{Introduction}
The positronium is a bound state of an electron-positron pair ($e^-$-$e^+$), that arises in the framework of Quantum Electrodynamics (QED). 
%A (very small) correction by weak interactions between is here neglected. 
As such, it plays an important role in fundamental physics for being the lightest `element' \cite{Adkins:2022omi,Bass:2019ibo}, as well as in medical physics, mainly due to its applications in positron emission tomography technique (PET) \cite{Adkins:2022omi,Bass:2023dmv,moskal2019nrp,Harpen:2003zz, moskalexvivo, moskalinvivo}. The ground state of the positronium, known under the name para-positronium (p-Ps), corresponds to a spin-singlet state 
%with anti-parallel orientations of the electron and positron spins
described by the spectroscopic quantum numbers $n$ $^{2S+1}L_J=$ $1$ $^1S_0$, 
or equivalently by using the relativistic notation as $J^{PC}=0^{-+}$, hence a pseudoscalar state just as the pion in QCD. 

The para-positronium is an unstable short-living state with a mean lifetime $\sim 0.125$ ns. It decays into photons due to the annihilation of its components. Charge-conjugation conservation requires that the decay of p-Ps occurs only into an even number of photons, with the  $\gamma \gamma$ mode (p-Ps$\rightarrow \gamma \gamma$) being by far the dominant one. 
This decay width can be expressed by (see e.g. Ref. \cite{Sen:2018wnx}): 
\begin{equation}
        \Gamma(Ps \rightarrow n\gamma)=\frac{1}{2J+1}\left|\psi(0)\right|^2\lim_{v\to 0}\left[4v\sigma (e^+e^- \rightarrow n \gamma) \right] 
        \text{ ,}
    \end{equation}
where $v$ and $\sigma$ are relative velocity and annihilation cross-section of $e^-$-$e^+$, 
 and where $\left|\psi(0)\right|^2 = m_e^2\alpha^3 /(8 \pi)$ is the spatial wave function at the origin (annihilation part of the amplitude). Namely, the ground-state wave function is 
$\psi(\vec{x})=(\pi a^3)^{-1/2}e^{-r/a}$, 
where $r = |\vec{x}|$ and $a = 2a_0$ (twice the Bohr radius of the hydrogen atom). 
At lowest order it becomes:
\begin{eqnarray}
    \Gamma\left(^{1}S_0 \rightarrow 2 \gamma\right)&=& \frac{1}{2} \frac{e^4 |\psi(\vec{x}=0)|^2}{\pi m^4} \int \limits_{0}^{\infty}|\vec{k}_1|^2\delta(2m-2|\vec{k}_1|)d|\vec{k}_1|= \nonumber \\
    &=& \frac{e^4|\psi(\vec{x}=0)|^2}{4\pi m^2}=\frac{4 \pi \alpha^2}{m^2}|\psi(\vec{x}=0)|^2 = \alpha^5 \frac{m_e}{2}  \text{ ,}
\end{eqnarray}
where the last result is the famous Wheeler-Pirenne formula \cite{wheeler,pirene} involving solely the fine-structure constant $\alpha$ and the electron mass $m_e$. 
In the framework of QED higher orders can be systematically evaluated, see e.g. Refs.  \cite{harris,adkins,czarnecki} and Table 1 for a summary. 
% Since many decades para-positronium is in the center of interest both for theoretical calculations and experimental observations. A significant achievement have been reached in both cases, especially for the decay rate p-Ps$\rightarrow \gamma \gamma$, that is the key point of this study. The corresponding analytical results described by QED as well as experimental result are summarized in Tab. \ref{tesummary}.      
\begin{table}[h!]
\renewcommand{\arraystretch}{1.5}
\centering
\begin{tabular}{|c|c|c|c|}
\multicolumn{4}{c}{$\Gamma$(p-Ps $\rightarrow 2 \gamma)$}\\
\hline \hline
\multicolumn{4}{c }{\textbf{Theory}} \\ \hline \hline
formula&result [$\mu s^{-1}$] &comment&ref.\\ \hline 
$\frac{\alpha^5 m_e}{2}$& $8032.5028(1)$ &lowest order (LO) & \cite{wheeler, pirene}\\
\hline
$\Gamma_0\left\{1+\frac{\alpha}{\pi}\left(\frac{\pi^2}{4}-5\right)\right\}$&$7985.249$ &LO+NLO& \cite{harris}\\
&&corrections & \\
\hline
$\Gamma_0\left\{1+\frac{\alpha}{\pi}\left(\frac{\pi^2}{4}-5\right)- \right.$&7989.6178(2)&LO+NLO+NNLO&\cite{adkins, czarnecki,Kniehl:2000dh,Abreu:2022vei}\\
$2 \alpha^2 ln \alpha +B_{2\gamma}\left(\frac{\alpha}{\pi}\right)^2-$&&corrections&\\
$\frac{3 \alpha^3}{2 \pi} ln^2 \alpha+C\frac{\alpha^3}{\pi}ln \alpha+$&&&\\
$\left.D\left(\frac{\alpha}{\pi}\right)^3\right\}$&&&\\
\hline
\multicolumn{4}{c }{\textbf{Experiment}} \\ \hline \hline
\multicolumn{3}{c| }{result $[\mu s^{-1}]$}& \multicolumn{1}{c }{ref.}\\ \hline
\multicolumn{3}{c|}{$7990.9(1.7)$ }& \cite{alram} \\
\hline
\end{tabular}
\caption{\label{tesummary} Summary of theoretical and experimental results for the decay rate p-Ps$\rightarrow 2\gamma$. The quantity $\Gamma_0=\alpha^5 m_e/2$ is the lowest-order (LO) result. The constants $A$, $B$, $C$ describe further corrections (NLO, etc.) and are reported in \cite{adkins}.}
\end{table}

Here, we intend to study the p-Ps within a composite model that makes use of the compositeness condition, originally proposed to describe the deuteron \cite{Weinberg, Hayashi}, in a way that resembles the treatment of QCD bound states \cite{Faessler:2003yf,Giacosa:2004ug,Giacosa:2007bs}. To this end, we extend the scalar model described in Ref. \cite{procsont}. The triangle diagram of Fig. 1 leading to $\gamma \gamma$ decay is calculated within this approach.  
Of course, the aim is not to go beyond the QED precision studies, but to learn how to deal with the nonperturbative vertex linking the positronium to its constituents \cite{Pestieau:2001ki}. In the following, a detailed discussion on this will be provided.

%definitely this would be not trivial. We rather try to find a connection between our approach and that of QED in order to see what we can learn about our treatment when compare to well-established results. On the other hand our goal is to inspect if our model can shed light on novel insights for the positronium described in non-relativistic picture. 

%In particular, in our model the two-body dacay of para-positronium into a pair of photons is described by the relativistic interaction Lagrangian that need to satisfy some specific conditions required for the bound state. Thus, in the following, we make use of the Weinberg compositeness condition, originally proposed to describe the deuteron \cite{Weinberg, Hayashi}. 

%We show that the inclusion of the triangle-shaped diagram into calculations reveals some main features of the process p-Ps$\rightarrow \gamma \gamma$. An interesting aspect is the positronium electron-positron vertex that seems to play a crucial role within our approach. In the following, a detailed discussion on this will be provided. 

\section{The composite model}
\begin{figure}[htb]
\centerline{%
\includegraphics[width=0.54\linewidth]{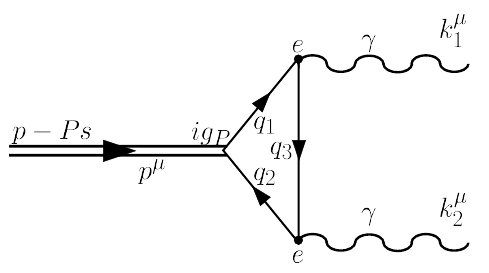}}
\caption{Triangle-shaped diagram for the process p-Ps$\rightarrow \gamma \gamma$ . }
\label{fig:trianglediagram}
\end{figure}

The Lagrangian of our model, describing the decay (and interaction) of para-positronium (with mass $M_P$) into two massless photons, reads:
\begin{equation}
\label{lag}
    \mathcal{L}_{int}=g_{P}P(x)\bar{\psi}(x)i\gamma ^{5}\psi (x)-eA_{\mu }(x)%
\bar{\psi}(x)\gamma ^{\mu }\psi (x)
\text{ ,}
\end{equation}
where $P(x)$ stands for the pseudoscalar positronium field, $\psi (x)$ is the electron field,  $A_{\mu}$ is the photon field, $e$ is the electric charge of the proton, and finally $g_P$ is the positronium-constituent coupling constant.  Note, there is no direct coupling between p-Ps and photons: the process p-Ps$\rightarrow \gamma \gamma$  is realized through triangle-shaped diagram with virtual electrons circulating in it, see  Fig. \ref{fig:trianglediagram}. 

At this point one needs to specify the kinematics of the two-body decay illustrated in Fig.  \ref{fig:trianglediagram}. For what concerns the external momenta one has: $p^{\mu}=(M_P, \vec{0})$, $k_1^{\mu}=(\omega,0,0, \omega)$, $k_2^{\mu}=(\omega,0,0, -\omega)$, with $\omega=\frac{M_P}{2}$, while the internal momenta are: $q_1=\frac{p}{2}+q$, $q_2=\frac{p}{2}-q$ and finally $q_3=\frac{p}{2}+q-k_1$. 

Let us now introduce one of the most important objects of our model, the triangle amplitude $I$ related to the diagram of the process depicted in Fig. \ref{fig:trianglediagram}, that reads: 
\begin{equation}
    I =i \int \frac{d^4q}{(2\pi)^4}\frac{\mathcal{F}(q,p)}{\left( q_{1}^{2}-m_{e}^{2}+i\varepsilon \right) \left(
q_{2}^{2}-m_{e}^{2}+i\varepsilon \right) \left( q_{3}^{2}-m_{e
}^{2}+i\varepsilon \right)} \text{ ,} \label{triangleampl}
\end{equation}
where $\mathcal{F}(q,p)$ is the `nonlocal' vertex function proportional to the Fourier transform  of  the wave function of the para-positronium. Note, this function is formally not present in Eq. (\ref{lag}), since the latter is local. It could be however easily included by rendering it nonlocal, see details in Refs. \cite{Faessler:2003yf,Giacosa:2004ug,Giacosa:2007bs,Wolkanowski:2015jtc}.  

The evaluation of the integral of Eq. (\ref{triangleampl}) has been done in two independent ways. The first one by using a Wick rotation, the second by making use of residue theorem. The basics of our formalism, including the discussion about the convergence of integral of Eq. (\ref{triangleampl}), has been presented in the QFT scalar toy model of Ref. \cite{procsont}. 

The triangle amplitude of Eq. (\ref{triangleampl}) can be written as:
\begin{equation}
    I=i \int \frac{d^{4}q}{(2\pi )^{4}}\frac{\mathcal{F}(q,p)}{D_1D_2D_3} \text{ ,}
\end{equation}
where:
\begin{eqnarray}
\label{d12}
&D_{1,2}&=\left(p/2 \pm q\right)^2-m_{e}^2+i \varepsilon=\left(M_P/2 \pm q^0\right)^2-\vec{q}^2-m_{e}^2+i \varepsilon, \nonumber \\ 
&D_3&=\left(M_P/2+q^0-k_1^0\right)^2-(\vec{q}-\vec{k_1})^2-m_{e}^2+i \varepsilon \text{ .} 
\end{eqnarray}
By setting $D_{1,2,3}=0$ one gets the corresponding poles:
\begin{equation}
\textrm{Poles of } D_1: \left\{ \begin{array}{l}
L_1=-\frac{M_P}{2}-\sqrt{\rho^2+q_z^2+m_{e}^2}+i\delta \\
R_1=-\frac{M_P}{2}+\sqrt{\rho^2+q_z^2+m_{e}^2}-i\delta
\end{array} \right. 
\text{ ,}
\end{equation}

\begin{equation}
\textrm{Poles of } D_2: \left\{ \begin{array}{l}
L_2=\frac{M_P}{2}-\sqrt{\rho^2+q_z^2+m_{e}^2}+i\delta \\
R_2=\frac{M_P}{2}+\sqrt{\rho^2+q_z^2+m_{e}^2}-i\delta
\end{array} \right.
\text{ ,}
\end{equation}

\begin{equation}
\textrm{Poles of } D_3: \left\{ \begin{array}{l}
L_3=-\sqrt{\rho^2+(q_z-k_z)^2+m_{e}^2}+i\delta \\
R_3=\sqrt{\rho^2+(q_z-k_z)^2+m_{e}^2}-i\delta
\end{array} \right.
\text{ .}
\end{equation}

The resulting decay width into $\gamma \gamma$ within our approach reads:
\begin{equation}
   \Gamma_{P \rightarrow \gamma \gamma}=\frac{1}{2} \frac{|\vec{k}_1|}{8 \pi M_P^2}2\left|8m_e4\pi \alpha g_P I  \frac{M_P^2}{4}\right|^2
   \label{neweq}
   \text{ ,}
\end{equation}
with $|\vec{k}_1|=\frac{M_P}{2}$.

\begin{figure}[htb]
\centerline{%
\includegraphics[width=0.4\linewidth]{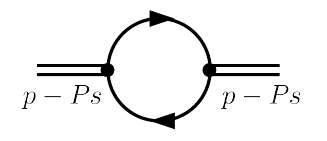}}
\caption{Loop diagram of the process p-Ps$\rightarrow e^-e^+ \rightarrow$ p-Ps.  }
\label{loop}
\end{figure}

The para-positronium is not an elementary particle, but an extended object emerging as a bound state of one electron and one positron. Thus, the positronium-constituent coupling constant $g_P$ entering the Lagrangian in Eq. (\ref{lag}) is not a free parameter of our model and it can be obtained by using the Weinberg's compositeness condition \cite{Weinberg,Hayashi}:
\begin{equation}
    g_P=\sqrt{\frac{1}{\Sigma'(s=M_p^2)}} \text{ with }
    \Sigma (s)=-i \int \frac{d^4q}{(2 \pi)^4} \frac{\mathcal{F}^2(\vec{q})^2}{D_1D_2}(-p^2/4+q^2-m_e^2)
    \text{ ,}
\end{equation}
where $\Sigma (s)$ is the loop function depicted in Fig. \ref{loop} as a self-energy  loop diagram.

Within our approach the vertex function $\mathcal{F}(q,p)$ turns out to be proportional to the wave function of the positronium in momentum coordinates, $A(\vec{q})$:
\begin{equation}
    \mathcal{F}(q,p)=\mathcal{F}(\vec{q}^2) \sim A(\vec{q})=\left(1+\frac{4}{\alpha^4+m_e^2}\vec{q}^2\right)^{-2}
    \text{ .}
    \label{corr}
\end{equation}
However, by simply setting it as equal ($ \mathcal{F}(q,p)=A(\vec{q})$) does not work because, for this naive Ansatz,  the value of the decay rate of the process p-Ps$\rightarrow \gamma \gamma$ turns out to be too small by a factor 2 when compared to experimental data.

Moreover, at first sight, it seems also that in Eq. (\ref{corr}) covariance is broken. That is not necessarily the case, since  we can interpret the vertex function as resulting from the Lorentz-invariant object \cite{Soltysiak}:
%when performing a deeper analysis, it turns out that the vertex function is not exactly equal to the momentum wave function of the positronium. Even if the necessary condition (agreement of analytical formula of our model with non-relativistic QED limit) is fulfilled, the value of the decay rate of the process p-Ps$\rightarrow \gamma \gamma$ is too small when compare to experimental data (factor 2). Still, the vertex function must behave as the wave function for small momenta. 
\begin{equation}
            \frac{-(pq)^2+p^2q^2}{p^2} = \vec{q}^2 \nonumber
            \text{ ,}
        \end{equation}
thus in the rest frame of the decaying particle, in which the four-momentum of the positronium is $p=(\sqrt{s},\vec{0})$, reduces to
        \begin{equation}
            \mathcal{F}(p,q)=\mathcal{F}\left(\frac{-(pq)^2+p^2q^2}{p^2}\right)=\mathcal{F}_{RF}(\vec{q}^2) 
            \text{ .}
        \end{equation}
As a consequence of this setup, in the rest frame there is no $q^0$ dependence (and no additional pole) resulting from the vertex function. 
%and for each pole the same $\mathcal{F}(\vec{q})$ appears. 

\begin{figure}[htb]
\centerline{%
\includegraphics[width=0.52\linewidth]{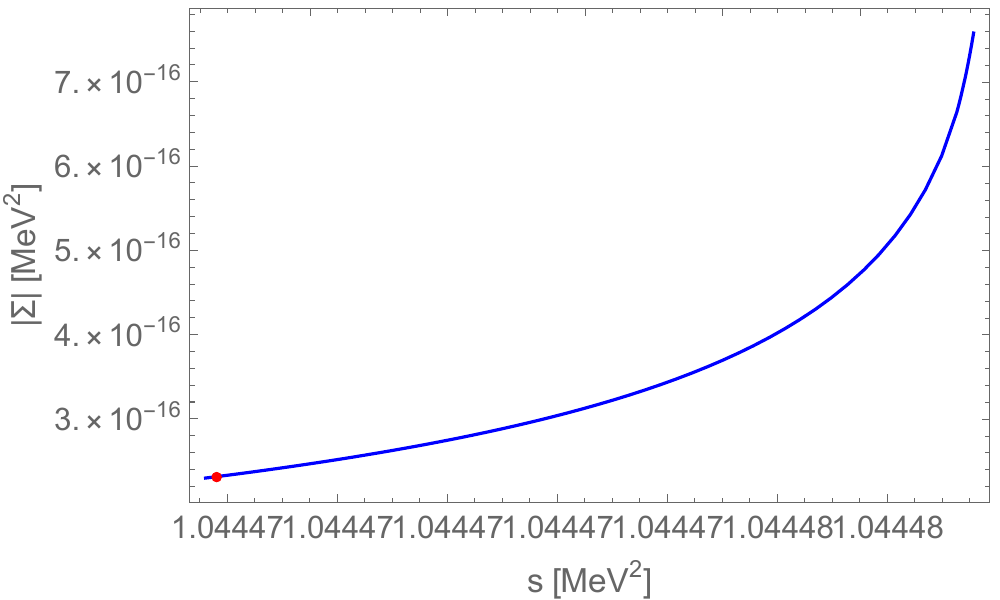}\includegraphics[width=0.5\linewidth]{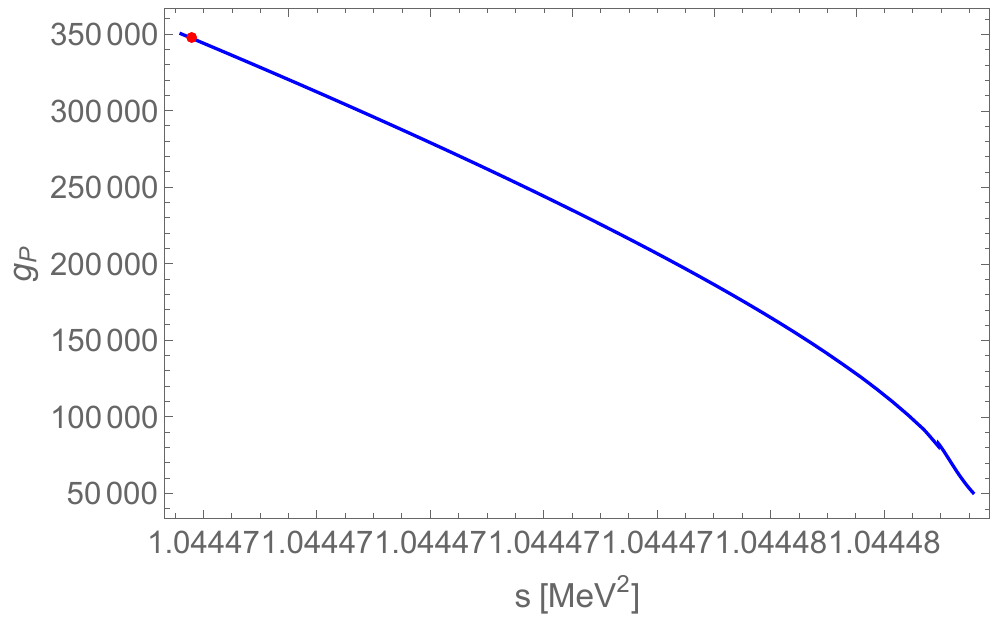}}
\caption{\label{sigmafunction} Left panel:  Dependence of loop function $\Sigma$ on the running square mass of the positronium $s=M_P^2$.  Right panel: dependence of coupling constant $g_{P}$ on $s=M_P^2$. The red dots represent the physical case.}
\label{fig:WRmodel3}
\end{figure}

Following Ref. \cite{Pestieau:2001ki} (see also the scalar model of Ref. \cite{procsont}) let us now consider another possibility for the vertex function, constructed as follow:
\begin{equation}
\label{vertfunction2}
    \mathcal{F}(\vec{q}^2)=\gamma^4 \left( \vec{q}^2+\gamma^2\right)^{-1}
\end{equation}
with $\gamma^2=m^2-\frac{M_P^2}{4}$.
First, in Fig. \ref{sigmafunction} we present the loop function $\Sigma(s=M_P^2)$ and the coupling constant for this choice. It is visible that for increasing the variable  $s$  the value of the loop function $\Sigma(s)$ also increases and reaches a cusp at the threshold located at the energy $2m_e$. Consequently, the coupling constant $g_P$ as a function of $s=M_P^2$ decreases for increasing $s$ and vanishes at the threshold.  

In Table \ref{polesresv2} we list the result of the decay width of the process p-Ps$\rightarrow \gamma \gamma$ obtained in our model by making use of the vertex function of Eq. (\ref{vertfunction2}). It is visible that the result involving the contribution of all three poles is now closer to the experimental value of 7990.9(1.7) $\mu s^{-1}$ than lowest-order the Wheeler-Pirenne result. In this respect, it can be shown that in the non-relativistic limit our model correctly reduces to $\alpha ^5 m_e/2$.

Moreover, it is interesting to study the role of each pole, where in Eq. (\ref{neweq}) we split $I = I_1 + I_2 +I_3$, where $I_k$ arises from the pole $R_k$. The contribution to the total decay rate from the first pole is by far the dominant one. Namely, the ratio of the amplitude contribution of the second pole w.r.t. the first one is $0.00170446$, while the contribution of the third w.r.t. first is $-0.00471415$. Interestingly, the third pole gives a negative contribution to the decay width.  This goes in the right direction but the contribution is even too strong, delivering a theoretical result smaller than the experiment. 
\begin{table}[h!]
\renewcommand{\arraystretch}{1.5}
\centering
\begin{tabular}{|c|c|}
pole(s) contribution &result [$\mu s^{-1}$]\\ \hline \hline
pole 1 & 7968.15\\ \hline
pole 1 + pole 2 & 7995.34\\ \hline
pole 1+pole 2+pole 3 & 7920.26\\ \hline
\end{tabular}
\caption{\label{polesresv2} Pole contribution to the $\Gamma$ (p-Ps $\rightarrow \gamma \gamma)$. }
\end{table}

Note, the result found in Ref. \cite{Pestieau:2001ki} for the same vertex function reads $7952.7 \mu s^{-1}$, the difference being due to the  Weinberg compositeness condition implemented in our approach. Still, we agree with the interpretation that QFT approaches implicitly contain the resummation of a certain class (but not all) of QED diagrams, see the discussion in Ref. \cite{Smith:2003sf}.

\section{Conclusions}
After a brief recall of the theoretical and experimental results on the para-positronium decay rate into two photons, we introduced a QFT composite model for its description. Upon evaluating the triangle diagram with virtual electrons and using the Weinberg compositeness condition, we calculated the $\gamma \gamma$ decay width. We have shown that the role of the vertex function is important. The choice of Ref. \cite{Pestieau:2001ki} delivers quite good results, but further improvement is needed for a determination agreeing with the very precise experimental result. 

In the future, one should test other choices for the vertex function, such as the promising covariant Ansatz of Ref. \cite{Gromes:1992ph}. In general the connection to the bound-state of quarks can also be useful \cite{Faessler:2003yf,Giacosa:2007bs,Smith:2003sf,Li:1990sx}.
Also, studies of the decay of radially excited states of para-positronium into photons as well as the decay of ortho-positronium into three photons represent an outlook of our model. 

\bigskip

\textbf{Acknowledgements:} This work was supported by the Minister of Science (Poland) under the `Regional Excellence Initiative' program (project no.: RID/SP/0015/2024/01). 

%uncomment the following lines to place a figure
%\begin{figure}[htb]
%\centerline{%
%\includegraphics[width=12.5cm]{Fig1}}
%\caption{Plot of ...}
%\label{Fig:F2H}
%\end{figure}

\end{document}